\documentclass[aps,pra,twocolumn,superscriptaddress]{revtex4-1}
\usepackage{times}
\usepackage{graphicx}
\usepackage{array}
\usepackage{amsmath}
\usepackage{amssymb}
\usepackage{mathrsfs}
\usepackage{color}
\usepackage{soul,xcolor}
\setstcolor{red}
\usepackage[utf8]{inputenc}
\usepackage[english]{babel}
\usepackage[utf8]{inputenc}
\usepackage[T1]{fontenc}    
\usepackage{amsmath,amsthm,amssymb,amsfonts, mathtools}
\usepackage{graphicx}
\usepackage{braket}

\begin{document}
\title{Temperature of the three-state quantum walk}

\author{Luísa Toledo Tude}
\affiliation{Instituto de F\'\i sica ``Gleb Wataghin'', Universidade Estadual de Campinas, Campinas, SP, Brazil}
\affiliation{School of Physics, Trinity  College  Dublin,  Ireland}
\author{Marcos C. de Oliveira }
\email{marcos@ifi.unicamp.br}
\affiliation{Instituto de F\'\i sica ``Gleb Wataghin'', Universidade Estadual de Campinas, Campinas, SP, Brazil}

\date{\today}

\begin{abstract}

Despite the coined quantum walk being a closed quantum system under a unitary evolution, its Hilbert space can be divided in two sub-spaces, which makes it possible for one to analyze one of the subsystems (the coin or the walker) as an open system in contact with a reservoir. In the present work we calculate the asymptotic reduced density matrix of the coin space of the three-state quantum walk in an infinite line, and use that result to analyze the entanglement between the chirality, and position space. We calculate the von Neumann entropy and the entanglement temperature per mean energy of the system in the asymptotic limit.
    
\end{abstract}


\maketitle

\section{Introduction}
Quantum Walks are a wide group of dynamical systems that represent the time evolution of a walker on a graph. Those are divided in discrete-time Quantum Walks and continuous time Quantum Walks. In this work we are interested in discrete-time Quantum Walks on a one-dimension position space, that is, on the lattice sites (vertices) of a line.
The study of Quantum Walks started as a generalization of the  classical random walks to quantum systems. However, some of their properties attracted the attention of researchers to the possibility of using them as a mathematical tool to build quantum algorithms. Among those properties we can cite: (i) The quantum walk in one dimension spreads ballistically, i.e, quadratically faster than the random walk. (ii) The amount of time taken to reach the limiting distribution of a quantum walk is quadratically faster than its classical counterpart. (iii) It has been proved that Quantum Walks, under particular conditions, can be used to implement a model of universal computation \cite{Universalcomp1,Universalcomp2}. (iv) The quantum walk can be used to simulate analogous systems, such as relativistic quantum mechanical systems \cite{articleQWandRelativisticQM}.

The three-state quantum walk is a generalization of the usual Hadamard Walk, where the probability of the walker to stay still (in the same vertex of the graph) in a time step of the dynamics of the system is also taken into account. In the classical case this additional consideration does not add much difference to the behavior of the system. On the other hand, for the quantum case, the addition of one degree of freedom in the chirality space causes a huge difference on the evolution of the position probability distribution. The probability amplitude of staying in the same vertex can generate a localization on the initial position. This property has been analyzed previously in \cite{3QW1,3QW3e4,3QWmatriz}. The Analysis the evolution of the partial Hilbert spaces in quantum walks has a close connection to the evolution of quantum open systems, when the walker Hilbert space acts as a reservoir for the  coin (chirality subspace), which tends to reach equilibrium in the asymptotic regime with the former. This dynamical process of equilibration clearly motivates one to question whether is it possible or not for it to be described as thermalization, in which case a definition of temperature for the walker subsystem must be defined. In fact, it was previously proposed \cite{Temperature,TGeneralizedQW,TLaws,Tinfo,vallejo2020temperature, TempAndres} that the entanglement between the position and chirality space allows one to define a so called "entanglement temperature", which in some specific situations may correspond to the Gibbs temperature. 

In the present work we analyze the entanglement between the position and chirality space in the asymptotic limit of the three-state Quantum Walk using the tools defined in Refs. \cite{Temperature,TGeneralizedQW,TLaws,Tinfo,vallejo2020temperature, TempAndres}. We have obtained an expression for the asymptotic coin density matrix as a function of the initial conditions. With that we calculated the entanglement entropy and derived a concept of temperature, showing that the system only achieves the Gibbs temperature in the asymptotic limit,  when it thermalizes with the bath. The paper is organized as follows. Firstly, in Sec. \ref{sec:3QW}, we introduce the notation making an overview of the three-state Quantum Walk on the infinite line. Then, in Sec. \ref{sec:Asymptotic}, we use the results of \cite{3QWmatriz} to calculate the asymptotic reduced density matrix of the three-state Quantum Walk. In Sec. \ref{sec:entropy}  we present our results concerning the thermodynamics of the three-state Quantum Walk, i.e, the values of entropy and temperature depending on the initial condition. Finally, in Sec. \ref{sec:C}, we present the last remarks and conclusions.

\section{Three-state quantum walk}\label{sec:3QW}
The principle behind the three-state quantum walk ---also known as the lazy quantum walk--- on a line is similar to the two-state one-dimension quantum walk \cite{PortugalBook}. The approach we employ to the analysis in this and in the following section is analogous to the one used to characterize the chirality space and thermodynamics of the two-state quantum walk \cite{Temperature}. 
\begin{figure}[ht!]
    \includegraphics[width=\linewidth]{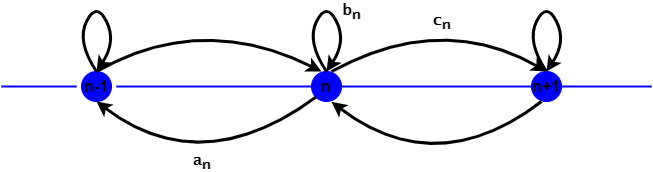}
\caption{Diagram of the three-state quantum walk. The coefficients $a_n$, $b_n$ and $c_n$ correspond to the left ($L$), no movement ($S$) and right($R$) chiralities, respectively.}
\label{fig:3QWdiagrama}
\end{figure}
The main difference between the two and three-state walks is that in the second case the chirality state space has three dimensions, therefore besides the possibilities of going to left or right, a probability of staying in the same site (vertex) is also taken into account. One can interpret this as a walk with a three-sided "coin". Fig. \ref{fig:3QWdiagrama} illustrates the possible steps of a walker that is at the $n$th site of the lattice.

Mathematically speaking, the joint system is composed by two subsystems, with Hilbert space given by
\begin{equation}
    \mathcal{H} = \mathcal{H}_C \otimes \mathcal{H}_P,
\end{equation}
where $\mathcal{H}_P$ is the position Hilbert space with infinite dimension, and $\mathcal{H}_C$ is the "coin" Hilbert space with dimension $3$. The state of the system, denoted as $\ket{\psi(t)}$, evolves according to a unitary evolution that can be recast in a sequence of two operators -one responsible for the flip of the "coin" ($C$) and the other for the conditioned shift of the walker in the sites of the lattice ($\text{Sh}$) 
\begin{equation}
    \ket{\psi(t + 1)} = U \ket{\psi(t)} = (\text{Sh} (C \otimes \mathbb{I}))\ket{\psi(t)}.
\end{equation}
The operator we will consider here to represent the action of the "coin" toss is
\begin{equation}
    C =  \frac{1}{3}   \begin{pmatrix}
-1&2&2\\
2&-1&2\\
2&2&-1
\end{pmatrix}, 
\end{equation}
known as Grover coin, spanned in terms of the  basis $\{|L\rangle,\,|S\rangle,\,|R\rangle\}$ representing, respectively, left, stay, and right chirality. The  shift on the walker, conditioned on the coin state is given by
\begin{equation}
    \begin{aligned}
        \text{Sh} =& \sum^{\infty}_{n = - \infty}\ket{n-1}\bra{n}\otimes \ket{L}\bra{L}\\
        &+\sum^{\infty}_{n = - \infty}\ket{n}\bra{n}\otimes \ket{S}\bra{S}\\
        &+\sum^{\infty}_{n = - \infty}\ket{n+1}\bra{n}\otimes \ket{R}\bra{R}.
    \end{aligned}
\end{equation}
The state of the system at time $t$ can be written as
 \begin{equation}
        \ket{\psi(t)} = U^t \ket{\psi_0} = \sum^{\infty}_{n = -\infty}
        \begin{pmatrix}
    a_n(t)\\
    b_n(t)\\
    c_n(t)
    \end{pmatrix} 
    \ket{n},
\end{equation}
 where
the action of the "coin" and shift operators can be summarized in the following recurrence relations
 \begin{equation}
     \begin{aligned}
         a_{n}(t+1) &=\frac{1}{3}(-a_{n+1}(t)  +2 b_{n+1}(t) + 2 c_{n+1}(t)),\\
         b_{n}(t+1) &= \frac{1}{3}(2 a_{n}(t)  - b_{n}(t) + 2 c_{n}(t)), \\
         c_{n}(t+1) &= \frac{1}{3}(2 a_{n-1}(t)  + 2 b_{n-1}(t) - c_{n-1}(t)).
     \end{aligned}\label{eq:evolucao3ab}
 \end{equation}
Therefore, defining the  Global chirality probabilities (GCP) of the walk in an analogous way to the two-state walk \cite{GCP}, 
 \begin{equation}
    \begin{aligned}
    P_L(t) =& \sum^{\infty}_{n = -\infty} |a_n (t)|^2, \\
    P_S(t) =& \sum^{\infty}_{n = -\infty} |b_n (t)|^2,\\
    P_R(t) =& \sum^{\infty}_{n = -\infty} |c_n (t)|^2.\label{eq:GCD3}
    \end{aligned}
 \end{equation}
one can use equation (\ref{eq:evolucao3ab}) to find the recurrence relation of the GCP
\begin{eqnarray}
        \begin{pmatrix}
    P_L (t+1)\\
    P_S (t+1)\\
    P_R (t+1)
    \end{pmatrix} 
    &=&\frac{1}{9}\begin{pmatrix}
    1 &4&4\\
    4&1&4\\
    4&4&1
    \end{pmatrix}
    \begin{pmatrix}
    P_L (t)\\
    P_S (t)\\
    P_R (t)
    \end{pmatrix}- \frac{\mathbb{R}[Q_1(t)]}{9}
    \begin{pmatrix}
    4\\
    4\\
    -8
    \end{pmatrix}\nonumber\\
     &&- \frac{\mathbb{R}[Q_2(t)]}{9}
    \begin{pmatrix}
    4\\
    -8\\
    4
    \end{pmatrix}
    - \frac{\mathbb{R}[Q_3(t)]}{9}
    \begin{pmatrix}
    8\\
    4\\
    4
    \end{pmatrix},
\end{eqnarray}
where the interference terms are given by
 \begin{equation}
     \begin{aligned}
     Q_1(t) &= \sum^{\infty}_{n = -\infty} a_n(t) b_n^{*}(t),\\
     Q_2(t) &= \sum^{\infty}_{n = -\infty} a_n(t) c_n^{*}(t),\\
     Q_3(t) &= \sum^{\infty}_{n = -\infty} b_n(t) c_n^{*}(t).
     \end{aligned}
 \end{equation}
 
\begin{figure}[ht!]
    \centering
    \includegraphics[width=\linewidth]{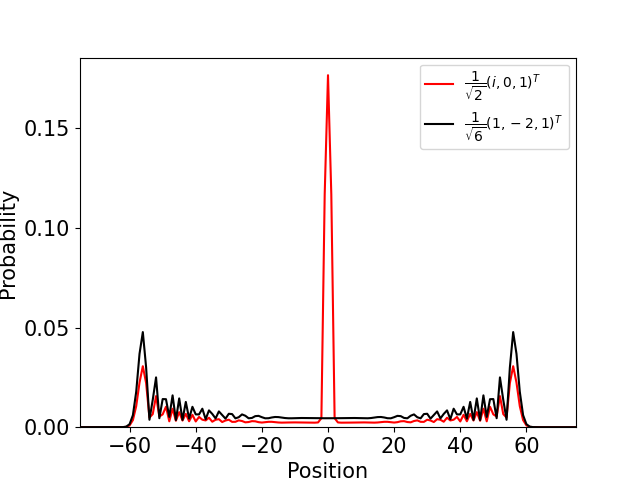}
\caption{(Color online) Distribution after $100$ time steps. Two different initial conditions were considered -- one that generates localization, $\ket{\psi_0} = \dfrac{1}{\sqrt{2}} (i,0,1)^T \ket{0}$, in light gray (red) and one that does not, $\ket{\psi_0} = \dfrac{1}{\sqrt{6}} (1,-2,1)^T \ket{0}$,in black.}
\label{fig:3QWpos}
\end{figure}

The simple unitary evolution of this quantum walk can be easily simulated. Fig. \ref{fig:3QWpos} shows the probability distribution of positions of the three-state quantum walk after $100$ time steps. We consider that the walker starts at position $0$. Depending on specific chirality initial conditions, a peak in the distribution is observed on the initial location of the walker, indicating localization. This happens due to the fact that one of the eigenvectors of the evolution operator of the Fourier space of the system is constant, as it will be further explored in the next section. 

To calculate the distributions presented in Fig. \ref{fig:3QWpos} we have to perform a partial trace over the chirality space. The aim of this work, however, is to analyze the entanglement between both --- chirality and position--- spaces and this can be done by tracing over any space, as the joint system state is pure. Since the chirality space has a lower dimension, choosing to work with its reduced density matrix  will make the calculations less laborious. The chirality reduced density matrix is given by
\begin{equation}
\rho_c (t) = \begin{pmatrix}
P_L(t)&Q_1 (t)&Q_2 (t)\\
Q_{1}^{*} (t) &P_S (t)& Q_3 (t)\\
Q_{2}^{*} (t)&Q_{3}^{*} (t)& P_R (t)
\end{pmatrix}.
\end{equation}

\section{Asymptotic limit}\label{sec:Asymptotic}
We are interested in equilibrium solutions of the evolution. Therefore we consider the asymptotic reduced state of the walk,
\begin{equation}
    \rho_{c,\infty}= \lim_{t \rightarrow \infty}\rho_c (t) .
\end{equation}
Throughout our calculations we consider a localized initial condition, which means that the chirality components can take any value, but the initial space component is known to be $\ket{0}$:
 \begin{equation}
        \ket{\psi_{0}^{0}} = 
        \begin{pmatrix}
    a\\
    b\\
    c
    \end{pmatrix} 
    \ket{0}.\label{eq:IC}
\end{equation}
A method similar to the one used in Ref. \cite{Fourier} to calculate the asymptotic state of the two-state quantum walk is applied - considering the Fourier transform of the wave function of the system, $ \Tilde{\Psi}(k,t)$, the equation describing the dynamics of the walk is 
\begin{equation}
    \Tilde{\Psi}(k,t+1) = \Tilde{M}^t \Tilde{\Psi}(k,0), 
\end{equation}
 where, in its diagonal form the matrix $\Tilde{M}$ has two time dependent eigenvalues ($\lambda_2$, $\lambda_3$) and a constant one ($\lambda_1 = 1$) \cite{3QW1,3QWmatriz}. The constant eigenvalue is responsible for the main difference in behavior of the two and three-state quantum walk, because it causes a localization around the walker's initial position. Therefore using its diagonal form, the evolution operator $\Tilde{M}$ can be decomposed as follows
 \begin{equation}
    \Tilde{M}^t = \Tilde{M}_{1} + \lambda_{2}^{t} \Tilde{M}_2 + \lambda_{3}^{t} \Tilde{M}_3.
 \end{equation}
The matrices $\Tilde{M}_{1}$, $\Tilde{M}_{2}$ and $\Tilde{M}_{3}$ are composed by the eigenvectors of $\Tilde{M}$. The state vector is obtained performing the inverse Fourier transform. A more detailed explanation of this procedure can be found on \cite{3QWmatriz}, where the asymptotic limit distribution was calculated with the matrices $U_1$, $U_2$ and $U_3$, defined on such a way that in the limit of $t\rightarrow \infty$ the state vector is
\begin{equation}
    \ket{\psi^{\infty}_{n}} = (U_1 + U_2 +U_3)\ket{\psi^{0}_{0}}.\label{eq:inftysate}
\end{equation}

To obtain the asymptotic reduced density matrix of the three-state quantum walk, we used the numerical calculations presented in the supplementary material of \cite{3QWmatriz}, where the matrices $\Tilde{M}_{1}$, $\Tilde{M}_2$ and $\Tilde{M}_3$ were obtained using the saddle point method. Those results suggest that to calculate the asymptotic density matrix, $\rho_{c,\infty} = \ket{\psi^{\infty}_{n}}\bra{\psi^{\infty}_{n}}$, the cross terms--- terms that involve different $U$ matrices --- should not be considered. Therefore
\begin{equation}
    \begin{aligned}
         \rho_{c,\infty} &= \ket{\psi^{\infty}_{n}} \bra{\psi^{\infty}_{n}} \approx U_1 \ket{\psi^{0}_{0}}\bra{\psi^{0}_{0}}U_1^{\dagger} \\& + U_2 \ket{\psi^{0}_{0}}\bra{\psi^{0}_{0}}U_2^{\dagger} +U_3 \ket{\psi^{0}_{0}}\bra{\psi^{0}_{0}}U_3^{\dagger}.
    \end{aligned}\label{eq:Assrho3QW}
\end{equation}
The analytical form of $\rho_{c,\infty}$ is too large to be displayed here. However, we present it in Appendix \ref{appendix}.

\begin{figure}[ht!]
    \centering
    \includegraphics[width=\linewidth]{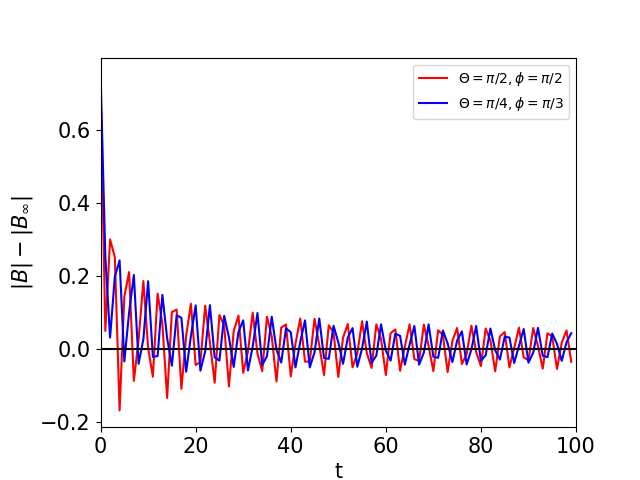}
    \caption{(Color online)Difference between the norm of the generalized Bloch vector calculated using the simulation and its asymptotic value. The initial conditions shown in the graph are $\ket{\psi^{0}_{0}} = (\cos{\theta},0,e^{i  \phi} \sin{\theta})^T \ket{0}$.}
    \label{fig:3QWBlochassintotico}
\end{figure}

In order to show that the asymptotic limit was calculated correctly, we compare the evolution in time of the norm of the generalized Bloch vector, $|{\bf B}|$, with the asymptotic value obtained by us, $|{\bf B_{\infty}}|$,  for different initial conditions --- Fig. \ref{fig:3QWBlochassintotico}.  The Bloch vector is obtained by the following relation, \cite{Bloch3QW},
\begin{equation}
    \rho_{c} = \frac{1}{3}(\mathbb{I}\text{d} + \sqrt{3}{\bf B \cdot \lambda}),\label{eq:Brho}
\end{equation}
where $\lambda = (\lambda_1,\lambda_2,\lambda_3,\lambda_4,\lambda_5,\lambda_6,\lambda_7,\lambda_8)$ are the Gell-Mann matrices \cite{Gell-Mann}. The results of Fig. \ref{fig:3QWBlochassintotico} suggest that the Bloch norm is indeed approaching the asymptotic value calculated using the asymptotic reduced density matrix and expression (\ref{eq:Brho}).

\section{Entanglement entropy and Temperature}\label{sec:entropy}
\begin{figure}[ht!]
    \centering
    \includegraphics[width=\linewidth]{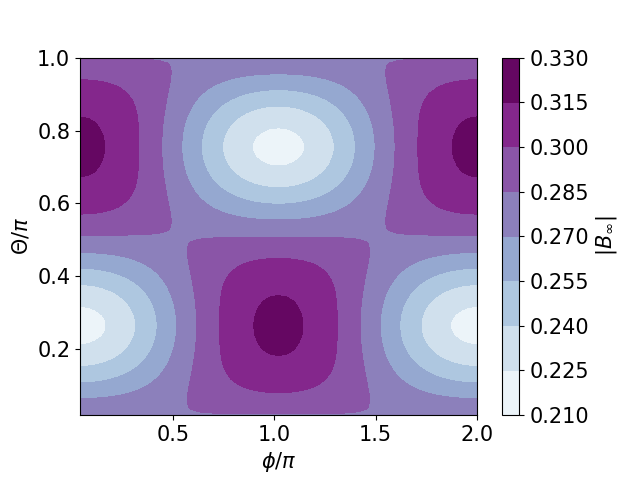}
    \caption{(Color online) Norm of the asymptotic generalized Bloch vector.}
    \label{fig:Bass}
\end{figure}

Fig. \ref{fig:Bass} shows values for the norm of the asymptotic generalized Bloch vector for all possible initial conditions of the type
\begin{equation}
     \ket{\psi^{0}_{0}} = 
    \begin{pmatrix}
\cos{\theta}\\
0\\
\sin{\theta} e^{i\phi}
\end{pmatrix} 
\ket{0}\label{eq:ci3QW}.
\end{equation}
We see that $|B_{\infty}|< 1$, meaning that although the initial chirality state is pure, as the evolution occurs, it goes to a mixed state. This behavior is due to the entanglement between position and the chirality spaces which can be quantified by the entanglement entropy, or von Neumann entropy
    \begin{equation}
        S =  -\text{Tr}(\rho_{c} \log \rho_{c}).\label{eq:VNentropy}
    \end{equation}

\begin{figure}[ht!]
  \includegraphics[width=.8\linewidth]{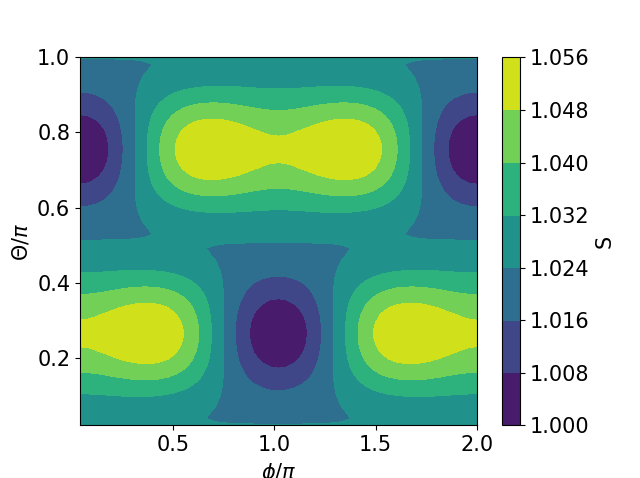}
  \includegraphics[width=.8\linewidth]{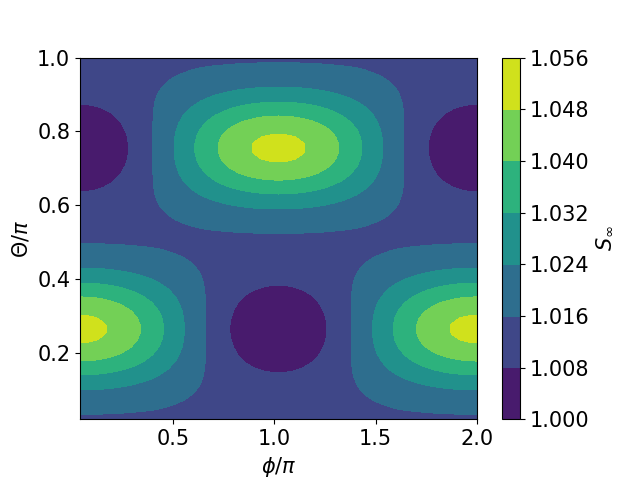}
\caption{(Color online) Entropy of the three-state quantum walks with initial condition of the type (\ref{eq:ci3QW}) after $t=100$ time steps (upper panel) and in the asymptotic limit (lower panel).}
\label{fig:S}
\end{figure}

The entanglement entropy can be calculated at any time step, since our goal here is to analyze the asymptotic limit of the three-state quantum walk, we present the entropy after 100 time steps and the asymptotic entropy of the walk for initial conditions of the type (\ref{eq:ci3QW}), without any loss of generality, Fig. \ref{fig:S}.

One important aspect to notice at this point is that the localization generated by the walk, i.e, the probability of the walker being at position $0$ in the asymptotic limit, is given by, \cite{3QWmatriz},
\begin{eqnarray}
       P(0,t=\infty)  &=& \begin{pmatrix}a^{*} & b^{*}& c^{*}\end{pmatrix} U_{1}^{\dagger} U_{1}\begin{pmatrix}a^{*} \\ b^{*}\\ c^{*}\end{pmatrix}\nonumber \\
        &=& (5-2\sqrt{6})\left[(2a + b)a^{*}+(a+b+c)b^{*}\right.\nonumber\\
        &&\left.+(b+2c)c^{*}\right],
\end{eqnarray}
where $a, b$ and $c$ are the initial chirality components. Therefore, all the initial conditions of the type (\ref{eq:ci3QW}) have localization $10 -4\sqrt{6}\approx 0.2$. This means that, although both features depend on the initial condition, there is no direct connection between the localization of the walk and its entanglement entropy.

A variable called entanglement temperature ($T_E$) can be defined with the help of the von Neumann entropy \cite{TempAndres}. The main idea  is to apply the definition from classical thermodynamics,
\begin{equation}
    \frac{1}{T} = \frac{\partial S_{\textit{class}}}{\partial E},\label{eq:Tclass}
\end{equation}
where $S_{\textit{class}}$ stands for the equilibrium thermodynamic entropy. Since the von Neumann entropy is the quantum extension of the statistical entropy - besides a $k_B$ factor - we extend this definition for the quantum case by considering the equilvalence of both. To simplify the calculation we consider $k_B = 1$. This definition was used in \cite{TempAndres} to calculate the entanglement temperature of the two-state discrete-time quantum walk, so, in principle we could also use it to calculate the entanglement temperature of the three-state quantum walk,
\begin{equation}
    \frac{1}{T_E} = \frac{\partial S}{\partial E}.\label{eq:defTE}
\end{equation}

The problem in calculating the entanglement temperature in the case of the three-state quantum walk is that the mean value of another observable, besides the energy, is required to be used as a constraint, \cite{vallejo2020temperature}. This means that, unlike the two-state case, we will not be able to calculate the entanglement temperature at an arbitrary  time. In the asymptotic limit, however, this restriction is surpassed and we see that since the system converges to an equilibrium distribution, the process can be analyzed as a thermalization and the asymptotic state is a Gibbs state, then we can calculate it. We also check that, in this case, the Gibbs temperature also respects the definition (\ref{eq:defTE}), meaning that our calculation is correct.

Let us start by writing the eigenvalues of the density matrix when the system (chirality) achieves equilibrium with the bath (walker)
\begin{equation}
    \tau_{j} = \frac{e^{-\beta \epsilon_j}}{Z}  = \frac{e^{-\beta \epsilon_j}}{e^{ -\beta \epsilon_1} + e^{ -\beta \epsilon_2}+e^{-\beta \epsilon_3}},\label{eq:taus}
\end{equation}
where the index $j$ stands for $1,2$ or $3$, and the Gibbs temperature per difference of energy can be obtained dividing any two of the three eigenvalues as
\begin{equation}
    \frac{T_G}{\epsilon_j' - \epsilon_{j}} = \frac{1}{\log{\left(\frac{\tau_j}{\tau_{j'}}\right)}}.\label{eq:gibbsT3QW}
\end{equation}
In the two-state quantum walk the temperature per difference of energy is well defined. In the present case, however, since there are three different ways of defining the temperature per difference of energy, it is more convenient to define the Gibbs temperature per mean energy. To derive this definition we set $H = {-{\bf v \cdot \lambda}}$, which is always possible since the set composed by the Gell-Mann matrices with the identity are a basis for the space of  $3\times3$ matrices and we will not take the identity  into account as it would only add a constant to the energy eigenvalues. Here ${\bf v}$ can be interpreted as a field or only mathematically as the coefficients of the linear combination of the Gell-Mann matrices, ${\bf \lambda}$.

\begin{figure}[ht!]
  \includegraphics[width=.8\linewidth]{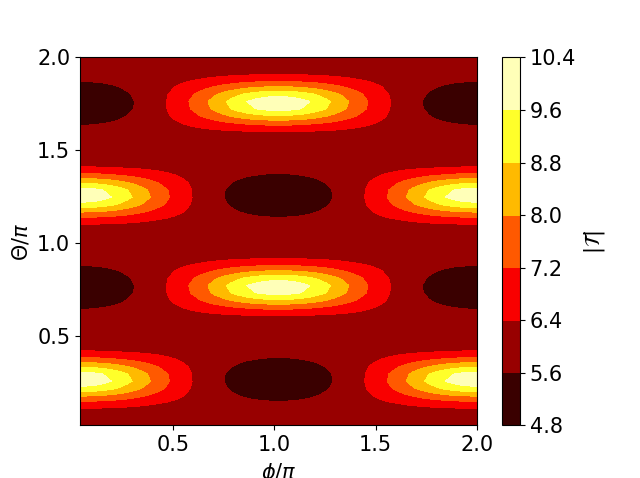}

  \includegraphics[width=.8\linewidth]{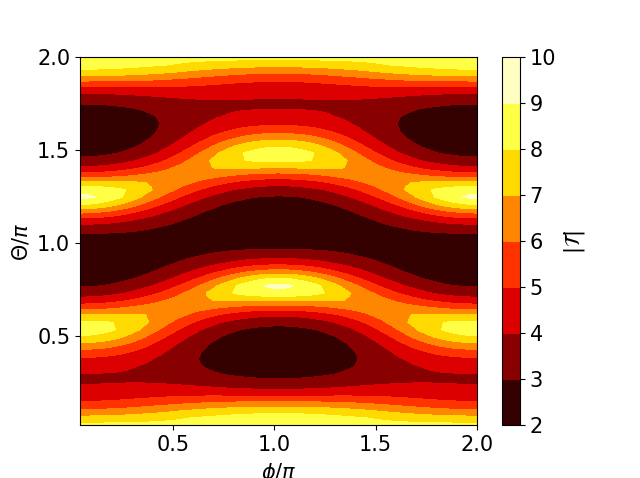}
\caption{(Color online) Color maps of the absolute value of the Gibbs temperature divided by the mean energy, for two initial conditions, $ \ket{\psi^{0}_{0_1}}$, in the upper panel, and $\ket{\psi^{0}_{0_2}}$, in the lower panel, from Eq. (\ref{eqci}).}
\label{fig:TgibbsMean}
\end{figure}

Using the fact that the mean energy is given by
\begin{equation}
    E = \text{Tr}[H \rho_s] = \epsilon_1 \tau_1 + \epsilon_2 \tau_2 + \epsilon_3 \tau_3,\label{eq:MeanE3qw}
\end{equation}
and that, since we chose to define $H$ as a traceless operator, the sum of the three eigenenergies is null,
\begin{equation}
    \epsilon_1 + \epsilon_2 + \epsilon_3 = 0.
\end{equation}
Dividing expressions (\ref{eq:taus}) of $\tau_1$ by the other two eigenvalues we find
\begin{equation}
    \begin{aligned}
        \frac{\tau_1}{\tau_2} &= e^{- \beta (\epsilon_1 - \epsilon_2)} = e^{- \beta (2 \epsilon_1 + \epsilon_3)}, \\
        \frac{\tau_1}{\tau_3} &= e^{- \beta (\epsilon_1 - \epsilon_3)},
    \end{aligned}
\end{equation}
where we used that $1 = \tau_1 + \tau_2 + \tau_3$ in the first line. Then, multiplying both expressions we get
\begin{equation}
    \frac{\tau_{1}^{2}}{\tau_2 \tau_3} = e^{- \beta (3 \epsilon_1)} \implies \epsilon_1 = - \frac{T_G}{3}  \log\left(\frac{\tau_{1}^{2}}{\tau_2 \tau_3}\right).
\end{equation}
Following an analogous procedure we obtain the expressions of the other eigenenergies
\begin{equation}
    \begin{aligned}
        \tau_2 \epsilon_2 &= - \frac{T_G}{3} \tau_2  \log\left(\frac{\tau_{2}^{2}}{\tau_1 \tau_3}\right),\\
        \tau_3 \epsilon_3 &= - \frac{T_G}{3} \tau_3  \log\left(\frac{\tau_{3}^{2}}{\tau_1 \tau_2}\right).
    \end{aligned}
\end{equation}
Now the mean energy (\ref{eq:MeanE3qw}) can be written as a function of the temperature and the eigenvalues of the density matrix only,
\begin{equation}
    \begin{aligned}
        E = &- \frac{T_G}{3}\left[\tau_1 \log\left(\frac{\tau_{1}^{2}}{\tau_2 \tau_3}\right) +\tau_2 \log\left(\frac{\tau_{2}^{2}}{\tau_1 \tau_3}\right) \right.\\& \left.+ \tau_3 \log\left(\frac{\tau_{3}^{2}}{\tau_2 \tau_1}\right)\right],
    \end{aligned}\label{eq:ET3qw}
\end{equation}
which leads to the final expression for the temperature per mean energy
\begin{equation}
    \begin{aligned}
        \mathcal{T} = \frac{T_G}{E} =& -3 \left[\tau_1 \log\left(\frac{\tau_{1}^{2}}{\tau_2 \tau_3}\right) +\tau_2 \log\left(\frac{\tau_{2}^{2}}{\tau_1 \tau_3}\right) \right.\\&\left.+ \tau_3 \log\left(\frac{\tau_{3}^{2}}{\tau_2 \tau_1}\right) \right]^{-1}.
    \end{aligned}
\end{equation}
Fig. \ref{fig:TgibbsMean} shows the result of $|\mathcal{T}|$ for two types of initial conditions, namely
\begin{equation}
    \ket{\psi^{0}_{0_1}} = 
    \begin{pmatrix}
\cos{\theta}\\
0\\
\sin{\theta} e^{i\phi}
\end{pmatrix} 
\ket{0},\,\, \text{and} 
\ket{\psi^{0}_{0_2}} = 
    \begin{pmatrix}
\cos{\theta}/\sqrt{2}\\
1/\sqrt{2}\\
\sin{\theta} e^{i\phi}/\sqrt{2}
\end{pmatrix} 
\ket{0}.\label{eqci}
\end{equation}

Manipulating expression (\ref{eq:ET3qw}) we can write $E$ as a function of the von Neumann Entropy and of the partition function, $Z = \text{Tr}[e^{- \beta H}]$. To do that we first isolate the energy
\begin{equation}
    \begin{aligned}
        E &= -\frac{T_G}{3}\left[\tau_1 \log(\tau_{1}^2) - \tau_1 \log(\tau_2 \tau_3) + \tau_2 \log(\tau_{2}^2)\right.\\&\left. - \tau_2 \log(\tau_1 \tau_3) + \tau_3 \log(\tau_{3}^2) - \tau_3 \log(\tau_1 \tau_2)
        \right],
    \end{aligned}
\end{equation}
then substituting the entropy,  $S = -\sum_{j=1}^{3} \tau_j \log(\tau_j)$, we find
\begin{equation}
    \begin{aligned}
        E &= -\frac{T_G}{3}\left[-2S - \tau_1 \log(\tau_2 \tau_3) - \tau_2 \log(\tau_1 \tau_3) \right.\\&\left. - (1 - \tau_2 - \tau_1) \log(\tau_1 \tau_2)\right],
    \end{aligned}
\end{equation}
which leads to
\begin{equation}
        E = -\frac{T_G}{3}\left[-3S  + \log\left(\frac{1}{\tau_1 \tau_2 \tau_3}\right) \right].
\end{equation}
Finally substituting the eigenvalues by the Gibbs probabilities we conclude that the von Neumann entropy is
\begin{equation}
    S = \frac{E}{T_G} + \log(Z)\label{eq:ESZ3qw}
\end{equation}
and, consequently, the definition (\ref{eq:defTE}) is valid for the Gibbs temperature.
\begin{equation}
    \frac{\partial S}{\partial E} = \frac{1}{T_G} \implies T_E = T_G.
\end{equation}
 This was already expected, since we started by using the Gibbs distribution to describe the system after it equilibrates with the bath, however it should be interpreted as a confirmation that our calculations are correct.

\section{Conclusions}\label{sec:C}
It is clear that the evolution of the chirality space of the two and three-state quantum walk is an equilibration process. Under certain conditions, \cite{equilibrium}, such a process can also describe thermalization. Here we analyzed subsystem of the coin of a three-state quantum walk from the point of view of the theory of quantum open systems. In references \cite{TempAndres,TGeneralizedQW} it was shown that for the two-state case the coin state converges to a thermal state. Here we considered that for the three-state quantum walk the equilibrium state is also thermal and used the canonical ensemble to derivate a definition of temperature per mean energy.

First we calculated the asymptotic reduced density matrix for the three-state quantum walk using the saddle point method. Then the von Neumann entropy was calculated for different initial conditions and different times, including the asymptotic limit. In opposition to the case of the Hadamard walk, the temperature of the three-state quantum walk can only be defined in the asymptotic limit. We calculated the expression for this temperature as a function of the initial conditions.

In our calculations we only assumed that the total energy of the system is zero and this does not imply any loss of generality because the differences between the eigenenergies are the relevant physical quantities, not their absolute values. Hence our expression for the temperature per mean energy extends to the general case of a three-state system that achieves equilibrium with the reservoir, not just the reduced state of the three-state quantum walk. This is in full agreement with the result by \cite{vallejo2020temperature}.
We also point to the fact that the relation obtained between the energy and entropy for the three-state quantum walk in the asymptotic limit, (\ref{eq:ESZ3qw}), is in accordance with the entropy-energy inequality discussed in Ref. \cite{EntropyEnergyInequality}.

Finally we would like to mention that after obtaining our results we became aware of two other papers that also analyze the limits of the three-state quantum walk, \cite{machida2013limit,machida2015limit}. Specifically, if one calculates the off-diagonal terms of the reduced density matrix using Lemma A.1 of \cite{machida2013limit} with $r=0$ the results are equal to the one we obtained in the Appendix. This is another confirmation that the asymptotic density matrix of the chirality space is correct.

\begin{acknowledgments}
The authors wish to thank Andrés Vallejo for helpful discussions. This work was financially supported by CNPq (Brazil).
\end{acknowledgments}

\appendix
\section{Asymptotic reduced density matrix}\label{appendix}
Here we present the asymptotic reduced density matrix of the three-state quantum walk calculated with Wolfram Mathematica, \cite{Mathematica}, with the help of the complement material of \cite{3QWmatriz}. This matrix was used to calculate the asymptotic results of the three-state quantum walk shown sections \ref{sec:Asymptotic} and \ref{sec:entropy}. 

Considering an initial condition of the type (\ref{eq:IC}) the reduced density matrix has the following form 
\begin{equation}
    \rho_{c,\infty} = \begin{pmatrix}
    \rho_{\infty,11}&\rho_{\infty,12}&\rho_{\infty,13}\\
    \rho_{\infty,21}&\rho_{\infty,22}&\rho_{\infty,23}\\
    \rho_{\infty,31}&\rho_{\infty,32}&\rho_{\infty,33}
     \end{pmatrix},
\end{equation}
where the matrix elements are presented in equation (\ref{eq:rhoass3QW}).
 \begin{widetext} 
\begin{equation}
\begin{aligned}
     \rho_{\infty,11} =&\frac{1}{48} (((48 - 11 \sqrt{6}) a + 6 (-8 + 3 \sqrt{6}) b + (-48 + 19 \sqrt{6})  c) a^{*} + 2 (3  (-8 + 3  \sqrt{6})  a + \sqrt{6}  (3  b + c))  b^{*}\\
    &+ ((-48 + 19  \sqrt{6}) a + \sqrt{6}(2 b + 5 c)) c^{*}) ,\\
    \rho_{\infty,22} =& \frac{1}{24}((3 \sqrt{6} a + 24  b - 10 \sqrt{6} b + 48  c - 19  \sqrt{6}  c) a^{*}+ 2  ((12 - 5  \sqrt{6})  a - 3  (-4 + \sqrt{6})  b + (12 - 5 \sqrt{6})  c) b^{*} \\
    &+ (48  a - 19 \sqrt{6}  a + 24  b - 10 \sqrt{6} b + 3 \sqrt{6} c)  c^{*}),\\
      \rho_{\infty,33} =& \frac{1}{48} ( (5 \sqrt{6} a + 2  \sqrt{6}  b - 48 c + 19  \sqrt{6}  c)  a^{*} + 2  ( \sqrt{6}  a + 3  \sqrt{6}  b - 24  c + 9  \sqrt{6}  c)  b^{*}\\
     &+ (  (-48 + 19  \sqrt{6})  a + 6  (-8 + 3  \sqrt{6})  b + ( 48 - 11 * \sqrt{6})  c)  c^{*}),\\
     \rho_{\infty,12} =( \rho_{\infty,21})^* =& \frac{1}{24}((3 (-8 + 3 \sqrt{6}) a + 96 b - 39 \sqrt{6} b + 144 c - 59  \sqrt{6} c) a^{*} \\
    &+ (3  \sqrt{6}  a + 24  b - 10  \sqrt{6} b + 48  c - 19  \sqrt{6}  c)  b^{*} + \sqrt{6} (a - b + c)c^{*}),\\
    \rho_{\infty,13} =( \rho_{\infty,31})^* =&\frac{1}{48}(((-48 + 19 \sqrt{6}) a + 288  b - 118  \sqrt{6}  b + 576  c - 235 \sqrt{6}  c)  a^{*}\\
    &+ 2  (\sqrt{6}  a + 48  b - 19  \sqrt{6}  b + 144  c - 59  \sqrt{6}  c) b^{*}+ (5  \sqrt{6}  a + 2  \sqrt{6}  b - 48  c + 19  \sqrt{6}  c)  c^{*}),\\
    \rho_{\infty,23} =( \rho_{\infty,32})^* =&  \frac{1}{24}( (\sqrt{6}  a + 48  b - 19  \sqrt{6}  b + 144  c - 59  \sqrt{6} c)  a^{*} - (\sqrt{6} a - 24b + 10 \sqrt{6}  b - 96  c + 39  \sqrt{6}  c)  b^{*}\\
    &+ (\sqrt{6}  a + 3  \sqrt{6}  b - 24  c + 9  \sqrt{6}  c)  c^{*}).
\end{aligned}\label{eq:rhoass3QW}
\end{equation}
\end{widetext}
 As expected, the matrix is Hermitian and $$\text{Tr}(\rho_{c,\infty})= |a|^2 + |b|^2 + |c|^2 = 1.$$

\bibliography{Bibliografia}   
\end{document}